\documentclass[runningheads]{llncs}
\usepackage[T1]{fontenc}
\usepackage{graphicx}
\usepackage{amsmath}
\usepackage{amssymb}
\usepackage{booktabs}

\usepackage{longtable}
\usepackage{booktabs}
\usepackage{array}

\usepackage{mathtools}
\usepackage{tikz}
\usetikzlibrary{positioning}
\usepackage{float}

\usepackage{bm}
\usepackage{algorithm}
\usepackage{algorithmic}

\usepackage[T1]{fontenc}
\usepackage{graphicx}
\usepackage[caption=false]{subfig}
\usepackage{enumitem}

\begin{document}
\title{Estimating Learners' Skill Acquisition \\ Without Temporal Information}
%

\author{Ryosuke Nagai\inst{1}\orcidID{0009-0005-2630-1842} \and
Kyohei Atarashi\inst{1}\orcidID{0009-0007-7765-8002} \and
Koh Takeuchi\inst{1}\orcidID{0000-0002-3245-888X} \and
Jill-Jênn Vie\inst{2}\orcidID{0000-0002-9304-2220} \and
Hisashi Kashima\inst{1}\orcidID{0000-0002-2770-0184}}
\authorrunning{R. Nagai et al.}
%
\institute{Kyoto University, Kyoto, Japan \\
\email{nagai.ryosuke@ml.kyoto-u.ac.jp}, \email{\{atarashi,takeuchi,kashima\}@i.kyoto-u.ac.jp}
\and
Inria, France \email{jill-jenn.vie@inria.fr}
}
\maketitle              

\begin{abstract}
Recent research in educational data mining, especially knowledge tracing, has focused on predicting learners' future knowledge states to support adaptive instruction. However, in many real-world educational settings, learning data are often available only as single-time-point assessments without temporal information, making existing time-series-based approaches difficult to apply.
In this paper, we propose a novel framework for predicting future skill acquisition using only snapshot data. Specifically, we address the problem of predicting the next skill to be acquired from skill mastery patterns estimated by cognitive diagnostic models (CDMs). In the absence of temporal information, we exploit inclusion relations among learners' skill sets to induce a pseudo-temporal ordering, interpreting expanding skill sets as a proxy for learning progression. To efficiently approximate unobserved acquisition paths, we introduce a neural model that captures latent skill acquisition dynamics through expected skill increments.
Experiments on both synthetic and real-world datasets demonstrate that the proposed method consistently outperforms baseline approaches, with particularly strong advantages as the skill space becomes larger. These results indicate that meaningful skill acquisition patterns can be inferred from snapshot data alone, providing a practical framework for adaptive learning support in data-constrained educational environments.
%
\keywords{
Educational Data Mining \and 
Next Skill Prediction \and 
Snapshot Data \and 
Cognitive Diagnostic Models \and 
Pseudo-temporal Ordering
}
\end{abstract}

\section{Introduction}
Understanding and predicting how learners develop knowledge over time is a central problem in educational data mining and learning analytics. Accurate prediction of learners' future knowledge can support a wide range of educational interventions, including personalized learning pathways, instructional planning, and early detection of learning difficulties. To address this problem, a substantial body of research, most notably knowledge tracing, has focused on modeling the evolution of latent knowledge states from time-stamped learner interaction logs~\cite{corbett1994knowledge,piech2015deep,shen2024survey}.
However, unlike classical knowledge tracing, many practical educational settings do not provide learner-wise temporal sequences, making it difficult to apply such models directly.


This limitation is particularly evident in large-scale and periodic assessment settings, where learner data are often collected as snapshots rather than as rich temporal traces. For example, international and national assessment programs such as PISA and NAEP evaluate learners at a particular stage of schooling through periodic assessments, and PISA explicitly characterizes each cycle as a cross-sectional study that provides a snapshot of students' developmental status~\cite{naep_about,pisa2022framework}. In addition, many large-scale assessments are administered in school-based, group-oriented formats that support population-level monitoring but do not directly provide learner-wise temporal logs~\cite{cresswell2015review}. 


Cognitive Diagnostic Models (CDMs) offer a widely used approach for representing learners' knowledge states from snapshot response data~\cite{junker2001cognitive,templin2010diagnostic,wang2024survey}. Based on a predefined mapping between items and skills, CDMs estimate learners' mastery of individual skills as binary skill mastery patterns. CDMs have also been applied to large-scale assessments; for example, prior work has used cognitive diagnostic assessment to infer learners' subskill profiles from PIRLS data~\cite{toprakyildiz2021pirls}. While CDMs are effective for static diagnosis, comparatively little attention has been paid to predicting future skill acquisition without temporal learning logs.

In this paper, we address this limitation by proposing a framework that predicts the next skill to be acquired using only snapshot-based skill mastery patterns. Our research question is: \textbf{Can learners' next skill acquisition be predicted solely from snapshot skill mastery patterns?} 

This capability has practical implications for settings in which only snapshot assessments are available. Estimating the next likely skill can help identify plausible next learning targets and support assessment-driven instructional decisions even in the absence of longitudinal logs.
Our contributions are twofold: (i) we formulate the problem of next-skill prediction based exclusively on CDM-derived snapshot representations, and (ii) we propose a pseudo-temporal transition model that learns skill-acquisition dynamics from snapshot inclusion patterns and predicts the next skill.

\section{Problem Setting}
\label{sec:problem_setting}
We consider the problem of predicting the next skill to be acquired by a learner based solely on their current skill mastery pattern. Let $K$ denote the total number of skills and $N$ the number of observed learners.

\paragraph{\bf{Observations}.}
For each learner $i$, the current skill mastery pattern is represented as a binary vector $\boldsymbol{s}_i \in \{0,1\}^K$, where $\boldsymbol{s}_i[k]=1$ indicates that skill $k$ has been mastered. The observed dataset is given by $\mathcal{D} = \{\boldsymbol{s}_i\}_{i=1}^{N}$.

\paragraph{\bf{Goal}.}
Given a learner's current mastery pattern $\boldsymbol{s}_i$, our goal is to estimate the conditional distribution of the next acquired skill, denoted by $P^{\ast}(\boldsymbol{y}_i \mid \boldsymbol{s}_i)$. Here, $\boldsymbol{y}_i = \boldsymbol{e}_{t_i} \in \{0,1\}^K$ is a one-hot vector corresponding to skill $t_i \in \{1, 2, \dots, K\}$, indicating the newly acquired skill.

\paragraph{\bf{Assumption (Non-forgetting).}}
We assume a non-forgetting learning process, in which once a skill is mastered, it remains mastered. Formally, if $\boldsymbol{s}_i[k]=1$ for skill $k$, the learner is assumed to retain that skill in all future states.

\section{Proposed Method}

\begin{figure}[t]
    \centering
    \subfloat[Skill acquisition paths.]{
        \includegraphics[width=0.5\linewidth,trim=8cm 6.5cm 6cm 4.5cm,clip]{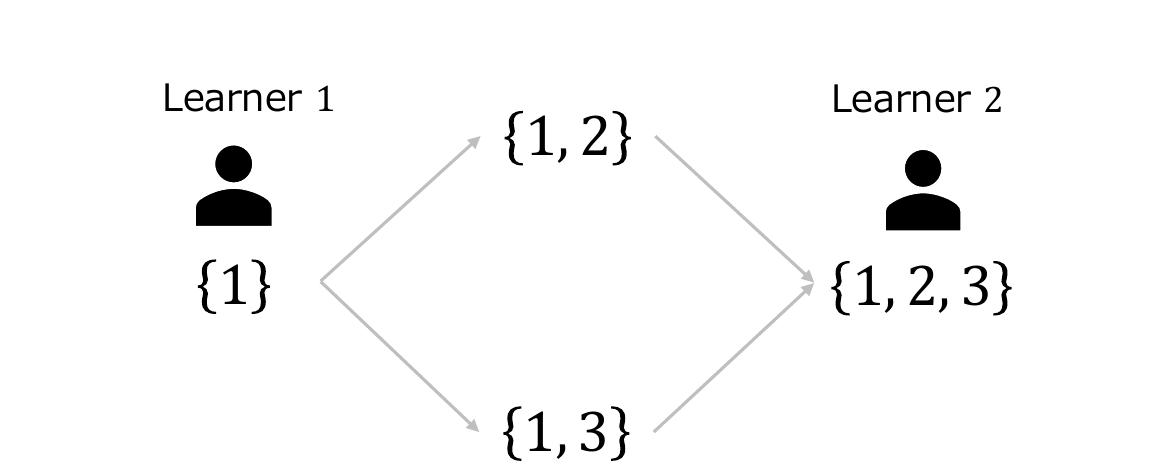}
        \label{fig:skill_acquisition_path}
    }
    \subfloat[Expected one-skill increment.]{
        \includegraphics[width=0.5\linewidth,trim=8cm 6.5cm 6cm 4.5cm,clip]{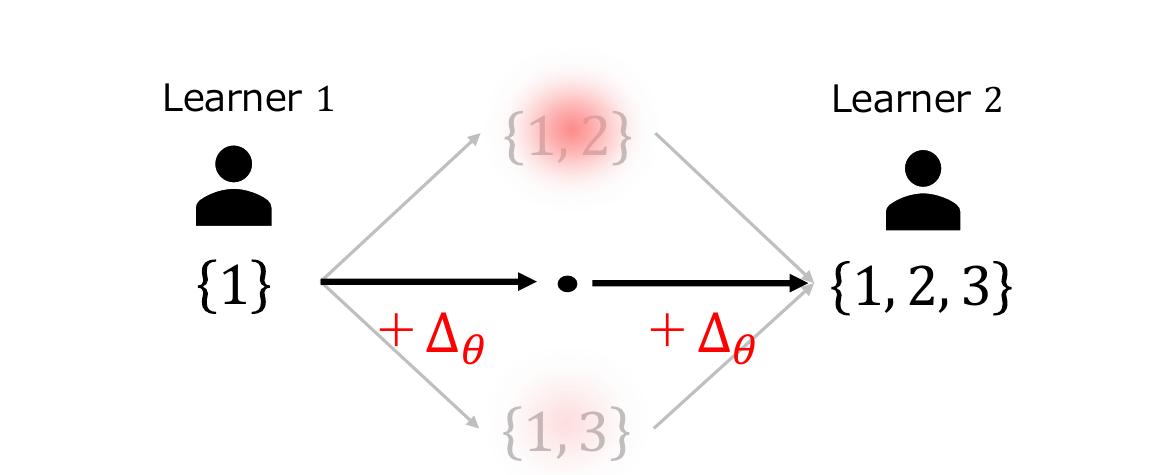}
        \label{fig:expected_one_skill_increment}
    }
    \caption{Illustration of path approximation via pseudo-temporal ordering.}
    \label{fig:path_approximation}
\end{figure}

\subsection{Basic Idea: Pseudo-Temporal Ordering}
The main challenge in our study is that no explicit temporal information is observed for each learner.
To address this, we focus on the diversity of skill mastery patterns in the observed data and construct a pseudo-temporal axis from these patterns.
This is particularly reasonable when plausible acquisition tendencies are broadly shared across learners, whether due to common skill dependencies or aggregate differences in skill difficulty.

We consider set-inclusion relations among mastered-skill sets across learners, where each learner's mastered-skill set consists of the skills that the learner has mastered. For example, in Fig.~\ref{fig:skill_acquisition_path}, learner~2's mastered-skill set is a superset of learner~1's. Under the non-forgetting assumption (Sec.~\ref{sec:problem_setting}), once a skill is mastered, it is never forgotten, and a learner's mastered-skill set expands monotonically over time. Therefore, a learner whose mastered-skill set is a superset of another learner's set can be viewed as one plausible later state. Such inclusion relations provide a useful structural signal for estimating latent skill acquisition dynamics from snapshot data.

In our pseudo-temporal view, acquiring one new skill constitutes the atomic step.
In practice, however, explicit skill acquisition paths are rarely observable from the data. Moreover, as the number of skills increases, the number of possible acquisition paths grows exponentially, making direct path enumeration infeasible. Therefore, instead of modeling complete acquisition paths, we introduce the expected increment of a single skill, denoted by $\Delta_\theta$, as the basic modeling unit (Fig.~\ref{fig:expected_one_skill_increment}). Since our target is the next acquired skill, this formulation provides a compact and task-aligned approximation of latent learning dynamics. Repeatedly accumulating these expected increments allows us to approximate intermediate states as continuous-valued representations.

\subsection{State Update Formulation via Expected Increments}
\begin{figure}[t]
    \centering
    \includegraphics[
    width=0.6\linewidth,
    trim=3cm 3.4cm 3cm 3.4cm,
    clip
    ]{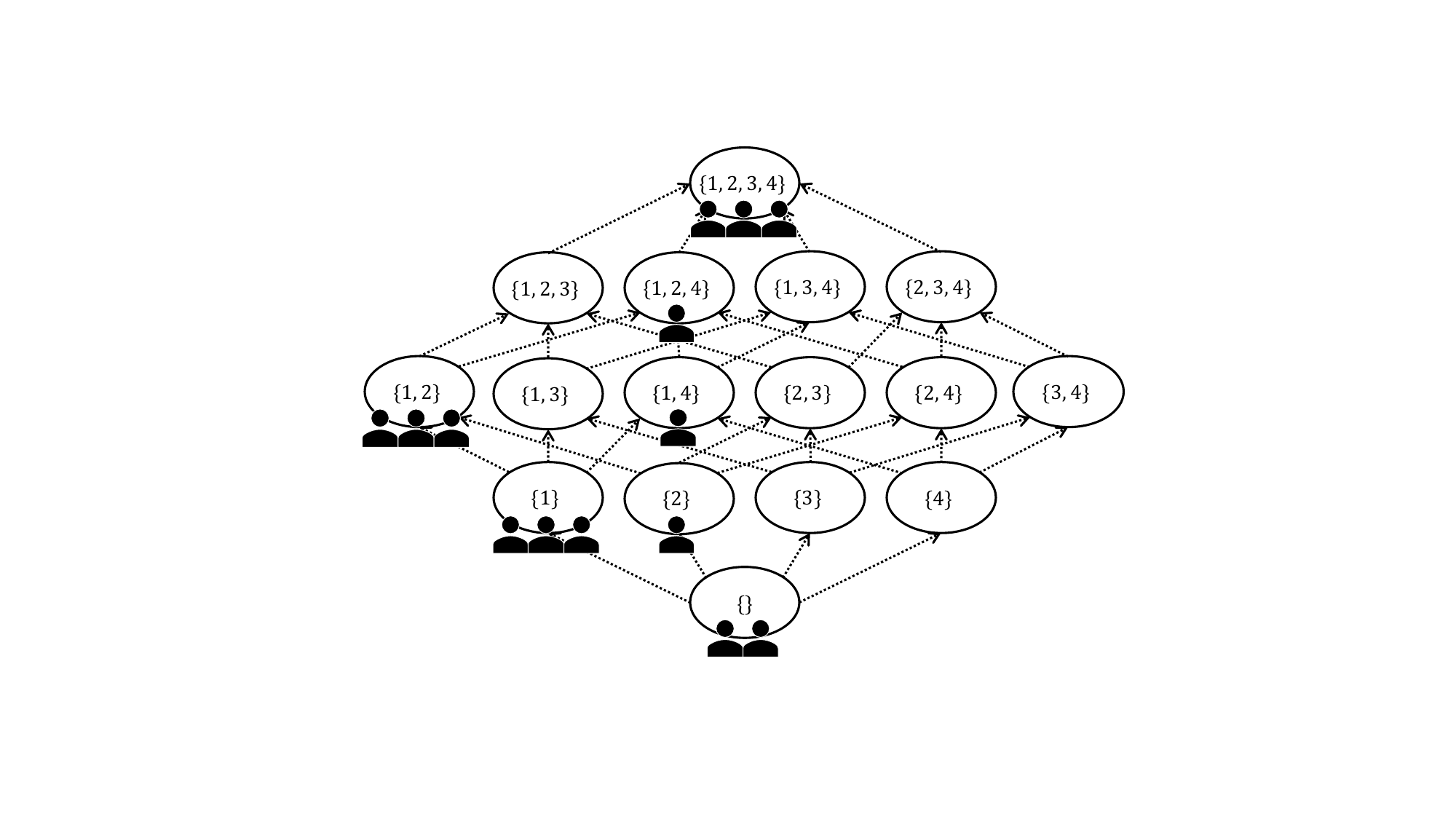}
    \caption{Skill mastery lattice for $K=4$, represented as a Hasse diagram. Each node denotes a mastery pattern, and each edge corresponds to the acquisition of one additional skill.}
    \label{fig:skill_mastery_hasse_diagram}
\end{figure}

Each observed binary mastery pattern can be viewed as a node in a skill lattice, illustrated as a Hasse diagram in Fig.~\ref{fig:skill_mastery_hasse_diagram}. In this representation, moving along an edge corresponds to acquiring exactly one new skill. Therefore, an observed state $\boldsymbol{s} \in \{0,1\}^K$ can be interpreted as the result of $\|\boldsymbol{s}\|_1$ successive one-skill acquisitions starting from the empty set.

Directly modeling all possible discrete acquisition paths is infeasible because their number grows rapidly with the number of skills. To obtain a tractable approximation, we introduce continuous intermediate states $\tilde{\boldsymbol{s}}^{(h)} \in [0,1]^K$ for $h \in \{0,\dots,K\}$, representing a relaxed mastery state after $h$ successive skill acquisitions. We then model learning as the accumulation of \emph{expected one-skill increments}. Formally, given an expected skill increment function $\Delta_{\theta}$ (defined in the next subsection), we define the state update as
\begin{align}
    \tilde{\boldsymbol{s}}^{(h+1)} \coloneqq \tilde{\boldsymbol{s}}^{(h)} + \Delta_{\theta}(\tilde{\boldsymbol{s}}^{(h)}), 
    \qquad
    \tilde{\boldsymbol{s}}^{(0)} = \mathbf{0}.
    \label{eq:skill_acquisition_process}
\end{align}

Starting from the empty state, we repeatedly add the expected next-skill increment to obtain a sequence of intermediate states. For learner $i$, we reconstruct the observed pattern $\boldsymbol{s}_i$ by applying this update $\|\boldsymbol{s}_i\|_1$ times, and define
$    \hat{\boldsymbol{s}}_i = \tilde{\boldsymbol{s}}_i^{(\|\boldsymbol{s}_i\|_1)}
$
as the predicted state corresponding to $\boldsymbol{s}_i$.

\subsection{Expected Skill Increment Model $\Delta_{\theta}$}
The expected skill increment function $\Delta_{\theta}$ maps a continuous mastery state to a distribution over the next skill to be acquired.
By Equation~\eqref{eq:skill_acquisition_process}, $\Delta_{\theta}$ should satisfy the following requirements: \textbf{(1) relaxed input}: it is defined on \emph{continuous} intermediate representations, and \textbf{(2) output constraint}: it assigns probability mass only to currently unmastered skills, i.e., $\Delta_{\theta}(\tilde{\boldsymbol{s}}) \in \Delta^K \cap [0, \bm{1}-\tilde{\boldsymbol{s}}]$, where $\Delta^K$ is the $K$-dimensional probability simplex.

To satisfy these conditions, we employ a feed-forward neural network with the Upper Bounded Softmax (UBSoftmax) function~\cite{atarashi2025box,martins2017learning}, which is an extension of Softmax with upper-bound constraints.
Formally, we define $\Delta_{\theta}$ as
\begin{align}
    \Delta_\theta(\tilde{\boldsymbol{s}}) &= \operatorname{UBSoftmax}(z(\tilde{\boldsymbol{s}}), \mathbf{1}-\tilde{\boldsymbol{s}}) \coloneqq \operatorname*{argmax}_{\boldsymbol{p} \in \Delta^{K} \cap [\boldsymbol{0}, \mathbf{1}-\tilde{\boldsymbol{s}}]} \boldsymbol{p}^\top z(\tilde{\boldsymbol{s}}) - \sum_{k=1}^K p_k \log p_k, \\
    z(\tilde{\boldsymbol{s}}) &= \boldsymbol{W}\tilde{\boldsymbol{s}} + \boldsymbol{b},
\end{align}
where $\boldsymbol{W}\in\mathbb{R}^{K\times K}$ and $\bm{b} \in \mathbb{R}^K$ are parameters to be trained.
By construction, $\Delta_\theta(\tilde{\boldsymbol{s}})$ lies in $\Delta^K \cap[\bm{0},\mathbf{1}-\tilde{\boldsymbol{s}}]$.

The model parameters $\theta =\{\bm{W}, \bm{b}\}$ are learned by minimizing
\begin{align*}
    \mathcal{L}(\theta) = \frac{1}{N}\sum_{i=1}^{N} \left\lVert
    \boldsymbol{s}_i - \tilde{\boldsymbol{s}}^{(\lVert \bm{s}_i\rVert_1)}_i \right\rVert_{2}^{2} +\lambda \lVert \theta \rVert_{1},
\end{align*}
where $\lambda \ge 0$ is the hyperparameter of regularization-strength.
The squared error term encourages accurate reconstruction of observed mastery patterns, while the $\ell_1$ regularization promotes sparse and interpretable dependencies among skills.

Through this training procedure, the model captures aggregate tendencies of skill set expansion along the induced pseudo-temporal axis.
\section{Experiments on Next-Skill Prediction}

\subsection{Experimental Setup}

\subsubsection{Evaluation Metrics.}
We evaluate the model's performance on the test set by predicting the subsequently acquired skills $\Delta \boldsymbol{s}_i$ from the current mastery pattern $\boldsymbol{s}_i$. Let $r = \|\Delta \boldsymbol{s}_i\|_1$ denote the number of newly acquired skills and $p_k$ be the predicted probability for skill $k$.

\begin{itemize}[leftmargin=*, itemsep=0pt, topsep=2pt]
    \item \textbf{New-Skill Accuracy (Acc.)}: Recall@r over unmastered skills.
    \item \textbf{Acquisition Cross-Entropy (ACE)}: $-\frac{1}{r} \sum_{k \in \Delta \boldsymbol{s}_i} \log p_k$, evaluating confidence on actually acquired skills.
    \item \textbf{Skill-State MSE (MSE)}: Mean squared error between predicted and true additional skill vectors.
\end{itemize}

For synthetic data, we additionally evaluate the reconstruction of next-skill probability distributions using \textbf{KL divergence (KL)} and \textbf{Jensen--Shannon divergence (JSD)}, weighted by the frequency of each mastery pattern.

\subsubsection{Baselines.}
We compare our approach with baselines applicable to prediction from single-time-point observations without learner-wise temporal logs: structure-based modeling (Bayesian Network), frequency-based prediction (Popularity), local transition estimation (Simple Markov), and an uninformed reference (Random).

\begin{itemize}[leftmargin=*, itemsep=0pt, topsep=2pt]
    \item \textbf{Bayesian Network}~\cite{plajner2016student,10.1142/S021848850400259X}: learns skill structures as a DAG using hill-climbing with BIC.
    \item \textbf{Popularity}: ranks skills by empirical acquisition frequency.
    \item \textbf{Simple Markov}: estimates next-skill probabilities from the observed frequency of next acquisition patterns given the current skill pattern.
    \item \textbf{Random}: assigns uniform probabilities to unmastered skills.
    \item \textbf{Oracle Upper Bound}: uses the same architecture as our model but is trained with true future skill labels, serving only as a reference upper bound.
\end{itemize}

All methods output a probability distribution over the next acquired skill. When multiple skills are observed as newly acquired, predictions are proportionally scaled to match the number of additions.

\subsection{Synthetic Data}
To evaluate performance under controlled learning dynamics, we generate synthetic datasets based on predefined skill prerequisite graphs. Learners acquire skills sequentially following these dependencies, with injected noise allowing occasional prerequisite violations. We evaluate 675 configurations (10 runs each), varying the number of learners, skills, noise levels, and acquisition tendencies. 
\subsubsection{Mean and Standard Deviation.}
As shown in Table~\ref{tab:metrics_summary}, across all settings, the proposed method consistently outperforms baseline approaches on all metrics except Acc. and closely approaches the oracle upper bound performance, indicating that meaningful signals of latent learning dynamics can be inferred from snapshot data.
\begin{table}[t]
\centering
\caption{Results on synthetic data (mean $\pm$ std).}
\label{tab:metrics_summary}
\resizebox{\textwidth}{!}{
\begin{tabular}{lccccc}
\toprule
Method & Acc. $\uparrow$ & ACE $\downarrow$ & MSE $\downarrow$ & KL $\downarrow$ & JSD $\downarrow$ \\
\midrule
Bayesian Network & 0.780 $\pm$ 0.103 & 0.500 $\pm$ 0.278 & 0.0774 $\pm$ 0.0426 & \underline{3.15 $\pm$ 3.04} & \underline{0.700 $\pm$ 0.592} \\
Popularity & \textbf{0.822 $\pm$ 0.089} & \underline{0.420 $\pm$ 0.248} & \underline{0.0580 $\pm$ 0.0278} & 3.66 $\pm$ 4.14 & 0.861 $\pm$ 0.795 \\
Simple Markov & 0.756 $\pm$ 0.124 & 1.466 $\pm$ 0.983 & 0.0778 $\pm$ 0.0395 & 34.6 $\pm$ 23.7 & 1.348 $\pm$ 0.977 \\
Random & 0.695 $\pm$ 0.136 & 0.492 $\pm$ 0.261 & 0.0750 $\pm$ 0.0354 & 5.70 $\pm$ 4.84 & 1.326 $\pm$ 1.110 \\
\midrule
\textbf{Proposed} & \underline{0.820 $\pm$ 0.091} & \textbf{0.397 $\pm$ 0.209} & \textbf{0.0571 $\pm$ 0.0288} & \textbf{2.68 $\pm$ 2.43} & \textbf{0.687 $\pm$ 0.639} \\
\midrule
Oracle Upper Bound & 0.832 $\pm$ 0.085 & 0.385 $\pm$ 0.203 & 0.0540 $\pm$ 0.0265 & 2.61 $\pm$ 2.47 & 0.673 $\pm$ 0.644 \\
\bottomrule
\end{tabular}
}
{\footnotesize Best and second-best results are shown in \textbf{bold} and \underline{underlined}, respectively.}
\end{table}

\subsubsection{Effect of Skill Size and Noise.}
\begin{figure}[t]
  \centering
  \includegraphics[width=4.5in, clip]{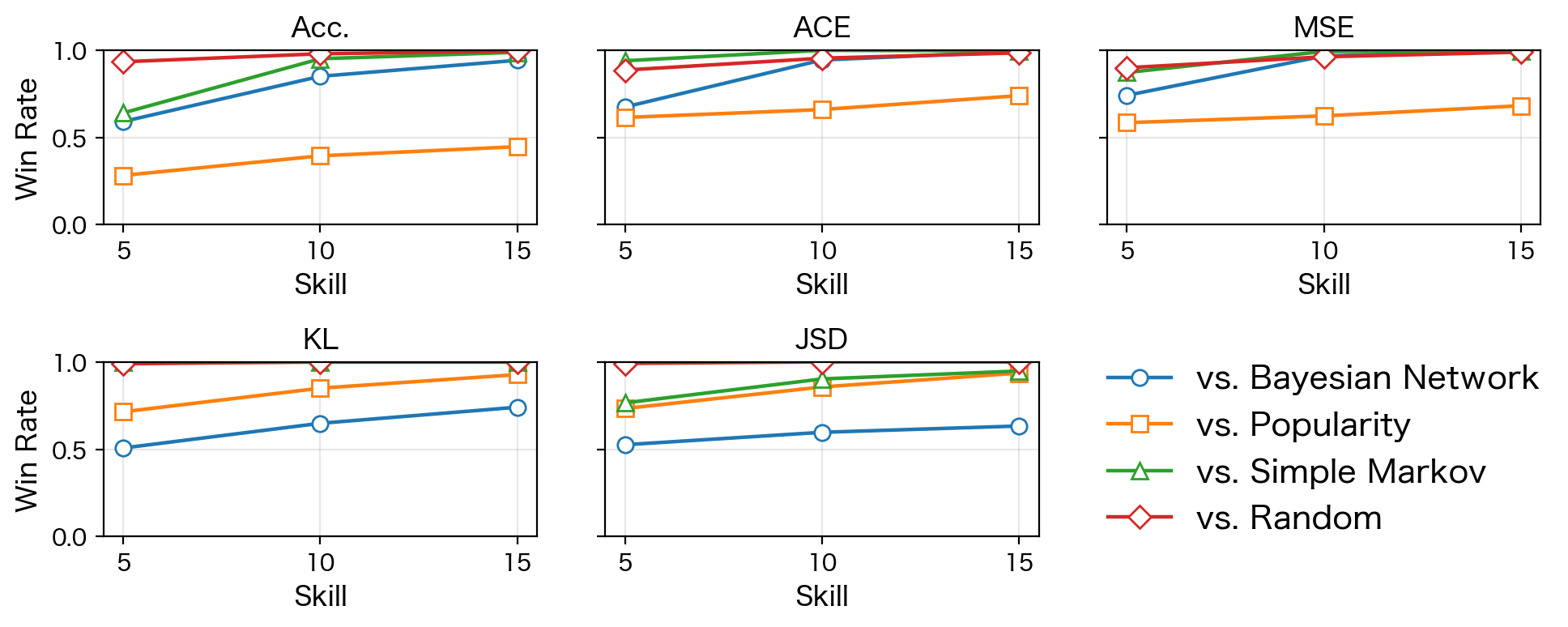}
  \caption{Win rate of the proposed method vs. number of skills (Synthetic Data).}
  \label{fig:skill_change}
\end{figure}

\begin{figure}[t]
  \centering
  \includegraphics[width=4.5in, clip]{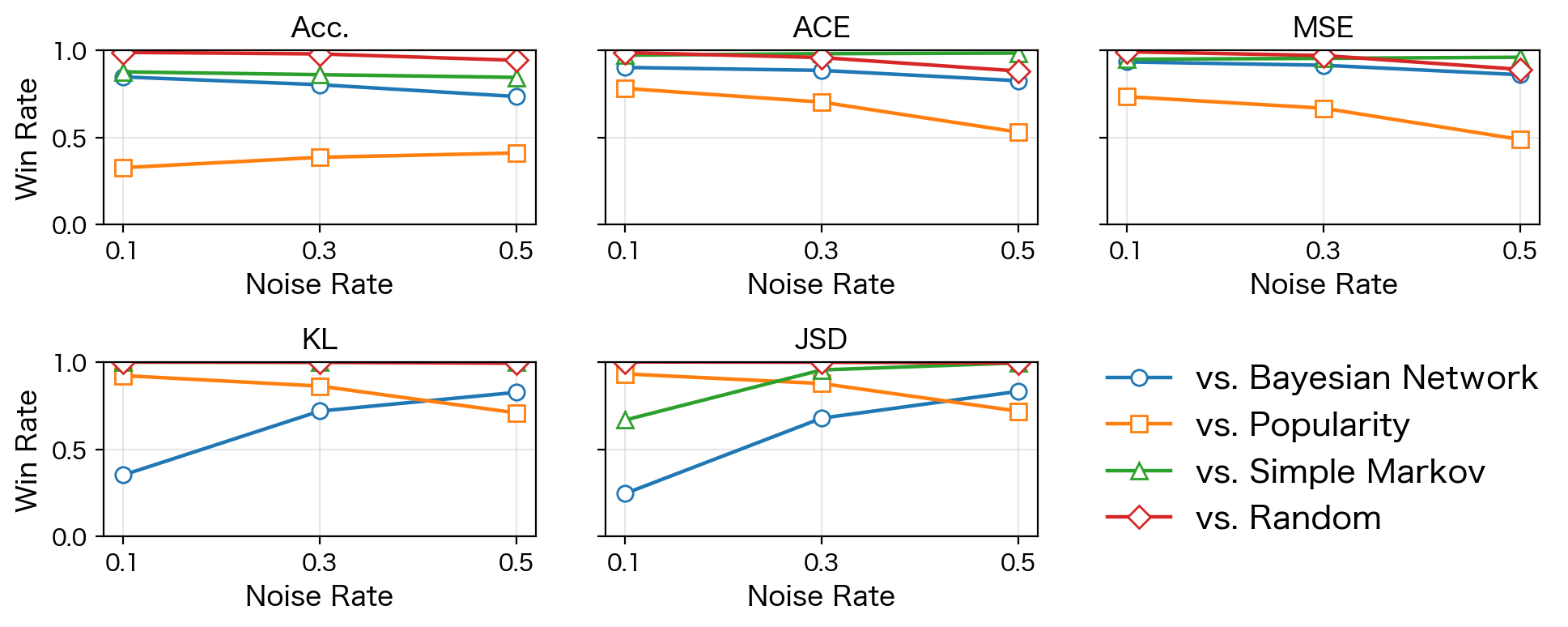}
  \caption{Win rate of the proposed method vs. noise rate (Synthetic Data).}
  \label{fig:noise_change}
\end{figure}

Figures~\ref{fig:skill_change} and~\ref{fig:noise_change} illustrate the win rates of the proposed method against each baseline across varying skill sizes and noise rates. The win rate represents the proportion of trials in which the proposed method outperformed the baseline.
As shown in these figures, the proposed method exhibits robust performance; in particular, its win rate consistently increases or remains high as the number of skills grows, indicating strong robustness to state-space expansion.
While the performance of several baselines degrades under higher noise rates, the proposed method maintains high win rates across all evaluation metrics, suggesting stable behavior even under noisy conditions.

\subsection{Real-World Data}
We further evaluate our method on the ASSISTments 2009--2010 dataset. To simulate a snapshot setting, we split each learner's interaction log into an early segment and the full log, and estimate current mastery from the early segment and future mastery from the full log using a DINA model~\cite{junker2001cognitive}. We selected 95 learners who attempted all 15 skills so that both current and future mastery patterns could be estimated consistently over a common skill space. 
This filtering yields a cleaner testbed for evaluation, but limits sample representativeness and should be relaxed in future work.
We performed 100 random train/test splits (70:30).

\subsubsection{Mean and Standard Deviation.}
As shown in Table~\ref{tab:15skill_results}, the proposed method achieves the best performance across all metrics on the real-world dataset, outperforming all baseline methods. These results suggest that the proposed approach can effectively predict future skill acquisition in this real-world benchmark setting.
\begin{table}[t]
\centering
\caption{Results on real-world data (15 skills, mean $\pm$ std.).}
\label{tab:15skill_results}
\resizebox{0.8\textwidth}{!}{
\begin{tabular}{lccc}
\toprule
Method & Acc. $\uparrow$ & ACE $\downarrow$ & MSE $\downarrow$ \\
\midrule
Bayesian Network & 0.648 $\pm$ 0.061$^{\dagger}$ & 0.853 $\pm$ 0.122$^{\dagger}$ & 0.0786 $\pm$ 0.0118$^{\dagger}$ \\
Popularity & \underline{0.760 $\pm$ 0.046}$^{\dagger}$ & \underline{0.645 $\pm$ 0.072}$^{\dagger}$ & \underline{0.0599 $\pm$ 0.0060}$^{\dagger}$ \\
Simple Markov & 0.581 $\pm$ 0.066$^{\dagger}$ & 1.811 $\pm$ 0.720$^{\dagger}$ & 0.0738 $\pm$ 0.0083$^{\dagger}$ \\
Random & 0.525 $\pm$ 0.065$^{\dagger}$ & 0.780 $\pm$ 0.090$^{\dagger}$ & 0.0696 $\pm$ 0.0056$^{\dagger}$ \\
\midrule
\textbf{Proposed} & \textbf{0.769 $\pm$ 0.052} & \textbf{0.626 $\pm$ 0.081} & \textbf{0.0571 $\pm$ 0.0063} \\
\midrule
Oracle Upper Bound 
& 0.807 $\pm$ 0.056
& 0.562 $\pm$ 0.075
& 0.0458 $\pm$ 0.0053 \\
\bottomrule
\end{tabular}
}
\par\smallskip
\parbox{0.8\textwidth}{\footnotesize $^{\dagger}$ indicates that the proposed method significantly outperforms the corresponding baseline (Wilcoxon signed-rank test, $p<0.05$).}
\end{table}

\subsubsection{Effect of Skill Size.}
\begin{figure}[t]
  \centering
  \includegraphics[width=4.7in, clip]{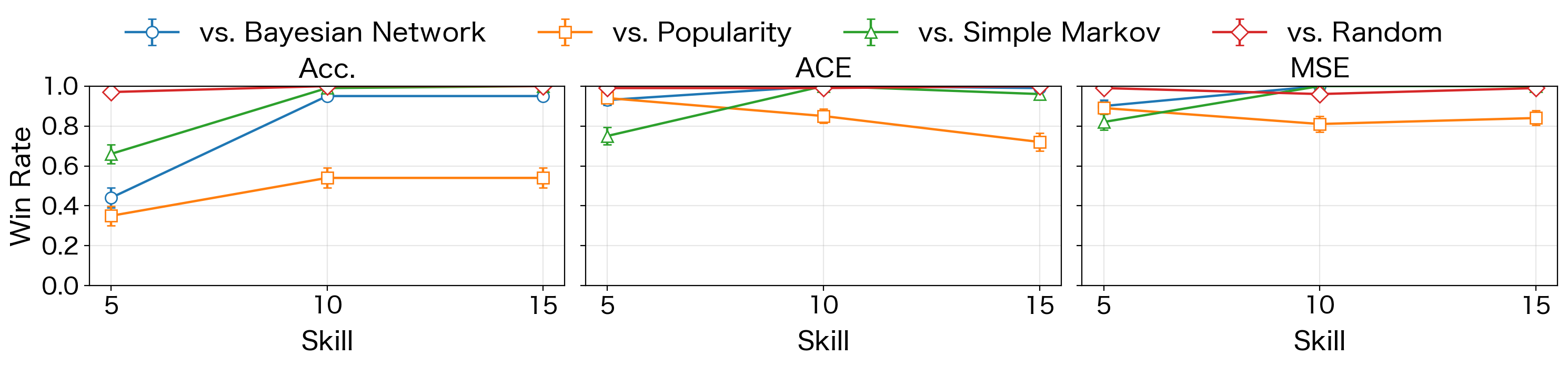}
  \caption{Win Rate of the proposed method vs. number of skills (Real-World Data).}
  \label{fig:skill_size}
\end{figure}

Consistent with the synthetic experiments, Figure~\ref{fig:skill_size} shows that the proposed method maintains high win rates against baselines as the number of skills increases. In particular, it consistently performs strongly on ACE and MSE, indicating better calibration and lower prediction error under higher-dimensional skill settings.

\section{Conclusion}
In this paper, we addressed the challenge of predicting learners' future skill acquisition in educational settings where temporal data is unavailable. We proposed a novel framework that infers latent learning dynamics solely from snapshot skill mastery patterns. By interpreting the inclusion relations among learners' skill sets as a pseudo-temporal ordering, we  modeled skill expansion as a proxy for learning progression.
Experiments on synthetic and real-world datasets demonstrate that our method consistently outperforms baselines. The results highlight the model's robustness to noise and skill space expansion, suggesting that meaningful acquisition patterns can be effectively inferred from snapshot data to support forward-looking instructional decisions.

Several limitations remain. Our current formulation relies on a non-forgetting assumption and is evaluated on a restricted real-world subset, so future work should examine more heterogeneous learning settings. An important next step is to incorporate forgetting mechanisms and further validate the framework on larger and more complex skill structures.

\begin{credits}


\end{credits}

\section*{Acknowledgment}
This work was supported by JST CREST Grant Number JPMJCR21D1.
\bibliographystyle{splncs04}
\bibliography{reference} 

\begin{thebibliography}{10}
\providecommand{\url}[1]{\texttt{#1}}
\providecommand{\urlprefix}{URL }
\providecommand{\doi}[1]{https://doi.org/#1}

\bibitem{atarashi2025box}
Atarashi, K., Oyama, S., Arai, H., Kashima, H.: Probability bounding: Post-hoc
  calibration via box-constrained softmax. arXiv preprint arXiv:2506.10572
  (2025)

\bibitem{corbett1994knowledge}
Corbett, A.T., Anderson, J.R.: Knowledge tracing: Modeling the acquisition of
  procedural knowledge. User Modeling and User-Adapted Interaction
  \textbf{4}(4),  253--278 (1994)

\bibitem{cresswell2015review}
Cresswell, J., Schwantner, U., Waters, C.: A Review of International
  Large-Scale Assessments in Education: Assessing Component Skills and
  Collecting Contextual Data. OECD Publishing and The World Bank (2015).
  \doi{10.1787/9789264248373-en}

\bibitem{junker2001cognitive}
Junker, B.W., Sijtsma, K.: Cognitive assessment models with few assumptions,
  and connections with nonparametric item response theory. Applied
  Psychological Measurement  \textbf{25}(3),  258--272 (2001)

\bibitem{martins2017learning}
Martins, A.F.T., Kreutzer, J.: Learning what’s easy: Fully differentiable
  neural easy-first taggers. In: Proceedings of the Conference on Empirical
  Methods in Natural Language Processing. pp. 349--362 (2017)

\bibitem{naep_about}
{National Center for Education Statistics}: About naep. National Assessment of
  Educational Progress (NAEP) website (2025), accessed: 2026-03-29

\bibitem{pisa2022framework}
{OECD}: PISA 2022 Assessment and Analytical Framework. OECD Publishing, Paris
  (2023). \doi{10.1787/dfe0bf9c-en}

\bibitem{piech2015deep}
Piech, C., Bassen, J., Huang, J., Ganguli, S., Sahami, M., Guibas, L.J.,
  Sohl-Dickstein, J.: Deep knowledge tracing. In: Advances in Neural
  Information Processing Systems. vol.~28 (2015)

\bibitem{plajner2016student}
Plajner, M., Vomlel, J.: Student skill models in adaptive testing. In:
  Proceedings of the Conference on Probabilistic Graphical Models. pp. 403--414
  (2016)

\bibitem{shen2024survey}
Shen, S., Liu, Q., Huang, Z., Zheng, Y., Yin, M., Wang, M., Chen, E.: A survey
  of knowledge tracing: Models, variants, and applications. IEEE Transactions
  on Learning Technologies  \textbf{17},  1858--1879 (2024)

\bibitem{templin2010diagnostic}
Templin, J., Henson, R.A.: Diagnostic Measurement: Theory, Methods, and
  Applications. Guilford Press (2010)

\bibitem{toprakyildiz2021pirls}
Toprak-Yildiz, T.E.: An international comparison using cognitive diagnostic
  assessment: Fourth graders' diagnostic profile of reading skills on pirls
  2016. Studies in Educational Evaluation  \textbf{70},  101057 (2021)

\bibitem{10.1142/S021848850400259X}
Vomlel, J.: Bayesian networks in educational testing. International Journal of
  Uncertainty, Fuzziness and Knowledge-Based Systems  \textbf{12}(1),  83--100
  (2004)

\bibitem{wang2024survey}
Wang, F., Gao, W., Liu, Q., Li, J., Zhao, G., Zhang, Z., Huang, Z., Zhu, M.,
  Wang, S., Tong, W.: A survey of models for cognitive diagnosis: New
  developments and future directions. arXiv preprint arXiv:2407.05458  (2024)

\end{thebibliography}

\end{document}